\documentclass[12pt,twoside,english,draftcls,peerreview,onecolumn,twoside]{IEEEtran}

\usepackage{cite}
\usepackage{amsmath,amssymb,amsfonts}
\usepackage{graphicx}
\usepackage{textcomp}
\usepackage{cases}
\usepackage{xcolor}
\usepackage{psfrag}
\usepackage{subfigure}
\usepackage{url}
\usepackage{stfloats}
\usepackage{amsmath}
\usepackage{array}
\usepackage{epsfig}
\usepackage{amssymb}
\newtheorem{theorem}{\textbf{Theorem}}
\newtheorem{lemma}{\textbf{Lemma}}
\newtheorem{remark}{\textbf{Remark}}
\newtheorem{proposition}{\textbf{Proposition}}
\usepackage{xcolor}
\usepackage{mathtools} 
\usepackage[ruled,vlined]{algorithm2e}

\begin{document}

\title{Performance Analysis of Uplink Adaptive NOMA Depending on Channel Knowledge}

\author{Mylene Pischella,~\IEEEmembership{Senior Member,~IEEE}, Ivan Stupia,~\IEEEmembership{Member,~IEEE} and Luc Vandendorpe,~\IEEEmembership{Fellow,~IEEE}
\tiny\thanks{Mylene Pischella is with Conservatoire National des Arts et M{\'e}tiers (CNAM), CEDRIC laboratory, Paris, France (contact: mylene.pischella@cnam.fr); Ivan Stupia and Luc Vandendorpe are with Institute of Information and Communication Technologies, Electronics and Applied Mathematics
Universit{\'e} catholique de Louvain, Louvain-la-Neuve, Belgium. This work was conducted while Mylene Pischella was a visiting researcher at Universit{\'e} catholique de Louvain and at ETIS UMR8051, CY University, ENSEA, CNRS, F-95000, Cergy, France.}\normalsize
}

\maketitle

\begin{abstract}
Non Orthogonal Multiple Access (NOMA) is a key technique to satisfy large users densities in future wireless networks. However, NOMA may provide poor performance compared to Orthogonal Multiple Access (OMA) due to inter-user interference.  In this paper, we obtain closed-form expressions of the uplink NOMA and OMA throughputs when no Channel State Information at Transmitter (CSIT) is available, and of the average data rates assuming that instantaneous rates should be larger than a minimum threshold when full CSIT is available. Analytical comparisons of OMA and NOMA prove that there is no global dominant strategy valid in all situations. Based on this conclusion, we propose  a new multiple-access (MA) strategy called NOMA-Adaptive (NOMA-A) that selects the best MA technique between OMA and NOMA. NOMA-A aims at maximizing the sum throughput in the no CSIT case, and   the probability that both users are  active in the full CSIT case. NOMA-A  is shown to outperform the other strategies in terms of sum throughput and rate.  \\
\vspace{0.1cm}
\end{abstract}

\section{Introduction}
Non Orthogonal Multiple Access (NOMA) is considered as one of the key techniques in fifth generation and beyond (B5G) technologies. Among the various proposals to multiplex several users on the same radio resources \cite{ChenVisoz18, DingBhargavaJSAC17}, power domain NOMA (PD-NOMA) \cite{DaiComMag15, DingPoorComMag17, LiuHanzoProceedings17, IslamDobreSurvey17} appears as one of the most effective due to its simplicity and ubiquity. PD-NOMA (called NOMA hereafter to simplify) consists in using superposition coding (SC) at the transmitters and successive interference cancellation (SIC) at the receivers. In the uplink, SIC should be performed by descending order of the received channel gains in order to maximize individual rates \cite{HossainIEEEAccess16}, whereas the opposite order should be used in the downlink.  NOMA may be used in uplink to  provide massive connectivity in  Internet-of-Things (IoT) networks \cite{LiangNOMA_WCom17,ShirDohlerJSAC17}, in which case grant-free NOMA protocols to handle collisions may be required \cite{AbbasVuceticTCom19, WangIoT19}. NOMA also enables  accomodating for large cellular users densities, whether these users are ultra reliable low latency (URLLC) subject to stringent quality
of service (QoS) delay constraints \cite{EEMaximization_Musavian_TWC15,Tradeoff_Yu_Musavian_TWC16, BelloChorti2020} or enhanced Mobile Broadband (eMBB) for which more classical scheduling strategies such as proportional fairness (PF) may be used \cite{NOMAPFVTC14, PFNOMAVTC16, PischellaNOMA2019WCL}. NOMA performances can be enhanced by optimizing  users clustering (that is, determining which subset of users' signals should be superimposed) \cite{SedaghatTWC18, HossainIEEEAccess16, CelikGlobecom17}, as well as optimizing the allocation of clusters onto radio resources such as subcarriers or Resource Blocks (RB) in multi-carrier systems, and optimizing power allocation \cite{FairPAOviedoTVT18, ZengPoor19,  LeiNOMA16, TweedIEEEAccess17, ZhaiAdmissionCULNOMA18}.

In this paper, we  focus on the analytical performance of two-users uplink NOMA  in two cases. In the first case, transmitters have no Channel State Information (no CSIT). Performances are then assessed in terms of outage probability and throughput. In the second case, full CSIT is available, but we assume that users only transmit if their instantaneous data rate exceeds a given minimum value. Average data rates then quantify users performances. Outage probabilities have been studied in NOMA with one-bit feedback, using the common outage probability metric \cite{XuSchonerOutage16}, under high Signal to Noise Ratio (SNR) approximations \cite{DingPoorSigProcLetters14}, or using an approximation of inter-user interference \cite{LiuOutageNOMA18}. However, to the best of our knowledge, closed-form expressions of the individual throughputs and average data rates in uplink NOMA valid whatever the SNR regime and with exact expressions of the interference have not been derived yet. 

The second major contribution of this paper is to determine adaptive multiple access (MA) strategies, where  users may either choose OMA or NOMA depending on which strategy is the most beneficial.  An adaptive strategy called NOMA-Relevant selecting between OMA and NOMA was first introduced in \cite{PischellaNOMA2019WCL}.  This strategy was shown to be more efficient than state-of-the-art ones \cite{HossainIEEEAccess16, CelikGlobecom17} in order to minimize latency. It has also been used to achieve stringent QoS delay constraints when maximizing the effective capacity in \cite{BelloVTCArxiv20}. NOMA-Relevant is however not applicable without full CSIT knowledge, and it assumes that users would transmit all the time, independently of their achieved data rates, which may lead to low instantaneous data rates and some waste of users battery.

The main contributions of this paper are the following:  
\begin{itemize}
    \item We analytically study the behavior of two-users uplink OMA and NOMA in two cases: when no Channel State Information is available at the transmitters (no CSIT) and when full CSIT is available.
    \item  OMA and NOMA are compared for both strong and weak users. We prove that the best MA strategy to maximize the studied metric (throughput or average data rate) depends on the user and on the average SNR regime. 
\item We then propose a new MA strategy called NOMA-Adaptive (NOMA-A) that adaptively selects between OMA or NOMA. In the no CSIT case,  NOMA-A maximizes the sum throughput, whereas in the full CSIT case, it maximizes the probability that both users are  active. The performance of OMA, NOMA and NOMA-A are   compared analytically and through simulations. Numerical results are also provided for more than $K=2$ users to show that the proposed algorithms are still effective when $K$ increases. 
\end{itemize}

The paper is organized as follows: Section \ref{SectionSystemModel} introduces the system model. Section \ref{SectionNoCSIT} focuses on the case when both transmitters have no CSIT, and Section \ref{SectionFullCSIT} on  the case with full CSIT but with a minimum admissible instantaneous data rate per user.  Section  \ref{SectionResults} provides numerical assessments of the NOMA algorithms. Finally, Section \ref{SectionConclusions} concludes the paper.

\section{System model \label{SectionSystemModel}}
We consider two-user uplink power-domain NOMA with Successive Interference Cancellation (SIC) at the Base Station (BS). Users fading $h_i$ follows an i.i.d Rayleigh distribution. We denote the received power from user $i\in \{1,2\}$ at the BS as $x_i=P_i|h_i|^2$ where $P_i$ is equal to the transmitted power divided by large-scale fading. For a given channel realization, users are indexed as follows: user $B$ is the one with the largest received SNR (and is therefore called the strong user) and user $A$ is the one with  the lowest received SNR (called weak user). Therefore,  constraint $x_B  \geq x_A$  holds. NOMA requires users to be ordered for efficient decoding. In the following, the same notations (users $A$ and $B$) are also used in OMA, so that a fair comparison can be performed between both MA strategies. 
$|h_i|^2$ has unit mean exponential distribution and  $x_i$ follows an exponential distribution with mean $P_i$ and parameter $\lambda_i = 1/P_i$. We assume that  $P_2 \geq P_1$. 

As $x_B = \max\{x_1,x_2\}$ and $x_A = \min\{x_1,x_2\}$, the  probability density function (pdf) of $X_A$ is:
\begin{align} \label{pdfMinExponentialva}
    f_{X_A}(x_A)= (\lambda_1 + \lambda_2) e^{-(\lambda_1 + \lambda_2)x_A}.
\end{align}
And the  pdf of $X_B$ is:
\begin{align} \label{pdfMaxExponentialva}
    f_{X_B}(x_B)=   \lambda_1 e^{-\lambda_1 x_B} +  \lambda_2 e^{-\lambda_2 x_B}-(\lambda_1 + \lambda_2) e^{-(\lambda_1 + \lambda_2)x_B}.
\end{align}
Finally, the joint pdf of $(X_A,X_B)$ is:
\begin{align} \label{pdfJointExponentialva}
    f_{X_A,X_B}(x_A,x_B)=   \lambda_1 \lambda_2 \left( e^{-(\lambda_1  x_A + \lambda_2 x_B)} + e^{-(\lambda_2 x_A +\lambda_1 x_B) }\right).
\end{align}
In order to simplify notations, $f_{X_A,X_B}(x_A,x_B)$ is denoted as $f(x_A,x_B)$ throughout the paper. 
Moreover,  $\rho = 1/(N_0)$ is defined as the inverse of  noise power.  In the following, it is referred to as the average SNR.

With NOMA, following the SIC process, the signal of the strong user $B$ is decoded first, while the interference from the weak user $A$ is considered as noise. Then if decoding enables to correctly recover the signal of user $B$, it is removed from the sum received signal, therefore enabling  the signal of  user $A$ to be decoded interference-free. Consequently, the instantaneous capacity of user $B$ in NOMA is:
\begin{align} \label{R2NOMADefinition}
    R_B  = \log_2\left(1+ \frac{\rho x_B }{1+\rho x_A }\right) 
\end{align}
and the instantaneous capacity of user $A$ in NOMA, assuming perfect decoding of user $B$, is:
\begin{align}\label{R1NOMADefinition}
    R_A  = \log_2\left(1+ \rho x_A \right).
\end{align}
With OMA, the instantaneous capacity of user $k\in \{A,B\}$ is:
\begin{align}\label{ROMA}
    \tilde{R}_k  = \frac{1}{2}\log_2\left(1+ 2 \rho x_k \right) 
\end{align}
where the $\frac{1}{2}$ coefficient is due to the fact that each user  transmits every other time slot, and the transmit power per time slot and user is twice that of NOMA, so that the total power budget over two time slots is the same as in NOMA. 

In the following, we consider two cases: either both transmitters have no CSIT, or they have full CSIT. Moreover, we assume that channel gains $(x_A, x_B)$ are unchanged during two consecutive time slots.

\section{Performance analysis and adaptive NOMA strategy in the no CSIT case \label{SectionNoCSIT}}

We first focus on a scenario with no CSIT but with channel knowledge at the receiver (CSIR) and we consider either the NOMA strategy or the OMA strategy. With both strategies, the same information quantity is sent over two consecutive time slots  and the same total power budget is also used. Let $\gamma$ be the minimum Signal to Interference Ratio (SINR) in NOMA, assumed equal for both users\footnote{We can notice that users could have different minimum SINR values, but that they should be set for users $\{1,2\}$ and not for the ordered users $\{A,B\}$, as the transmitters are not aware of whether they are the strong or the weak user. Considering different values of $\gamma_1$ and $\gamma_2$ leads to a very complex statistical problem that is not addressed in this paper. Moreover, setting the same minimum SINR value is more logical from an operational viewpoint, as this minimum SINR corresponds to a given maximum Bit Error Rate requirement for a specific modulation.}.    The information quantity sent by user $k$ in both time slots is then equal to $\log_2\left(1+ \gamma \right)$. We assume that $\gamma \geq 1$, so that the SIC decoding order in NOMA can only be user $B$ before user $A$.  Moreover, when $\gamma \geq 1$,  considering interference as noise at both users always leads to setting user $A$ in outage. Therefore, using SIC is mandatory. 

The information quantity sent by user $k$ in OMA in the only time slot where user $k$ is active over two consecutive time slots must be twice that sent in NOMA per time slot, and is given by:
\begin{align}
   \log_2\left(1+ \tilde{\gamma} \right)  = 2 \log_2\left(1+ \gamma \right). 
\end{align}
Consequently, the  minimum SNR in OMA $\tilde{\gamma}$ is defined as:
\begin{align} \label{EquivalentOMASNR}
   \tilde{\gamma} = 2 \gamma  + \gamma^2.
\end{align}

The amount of correctly decoded information is called the throughput. It is equal to the sent information quantity times the probability that this information quantity is correctly decoded, which is defined as $1$ minus the outage probability. Users are in outage if their instantaneous SINR is lower than $\gamma$. Let     $\phi_{k,N}(\rho)$ for  $k\in \{A,B\}$ be the probability that user $k$ is not in outage with NOMA. For user  $B$, it is given by:
\begin{align} \label{DefinitionPhi2N}
    \phi_{B,N}(\rho)& = P\left(x_B \geq \frac{\gamma}{ \rho } (1+\rho  x_A ) \right).
\end{align}

User $A$ is not in outage with NOMA if the BS has correctly decoded the signal of user $B$ using SIC and if the instantaneous SNR of user $A$ is larger than $\gamma$. This implies that whenever user $B$ is in outage, user $A$ is also in outage. Consequently, the probability that user $A$ is not in outage with NOMA is given by:
\begin{align} \label{DefinitionPhi1N}
    \phi_{A,N}(\rho) = P\left(\left(x_B \geq \frac{\gamma}{ \rho } (1+\rho  x_A )\right)  \cap \left( x_A \geq \frac{\gamma}{ \rho})\right) \right).
\end{align}
On the contrary, in OMA, the probability not to be in outage is independent for each user $k\in \{A,B\}$ and is given by: 
\begin{align} \label{DefinitionPhikO}
    \phi_{k,O}(\rho) = P\left( x_k \geq \frac{ \tilde{\gamma}}{2  \rho}\right).
\end{align}
Then, the throughput with NOMA for $k\in \{A,B\}$ is:
      \begin{align} \label{ThroughputNOMA}
        T_{k,N}  =  \phi_{k,N}  \log_2\left(1+\gamma\right).
            \end{align}
And similarly, the throughput with OMA for $k\in \{A,B\}$ is equal to:
      \begin{align} \label{ThroughputOMA}
        T_{k,O}  =  \phi_{k,O}   \log_2\left(1+\gamma\right).
    \end{align}

\subsection{Evaluation of  NOMA and OMA throughputs}

\subsubsection{NOMA probabilities}
For the strong user $B$, $\phi_{B,N}(\rho)$ depends on which constraint $x_B \geq x_A$ and $x_B \geq \frac{\gamma}{ \rho } (1+\rho  x_A )$ is the most stringent. As we assumed that $\gamma \geq 1$, $\frac{\gamma}{ \rho } (1+\rho  x_A ) \geq x_A$ always stands and the analytical expression of $\phi_{B,N}(\rho)$ is equal to:
\begin{align}\label{Phi2NCase1}
\phi_{B,N}(\rho)& = \int_{x_A=0}^{\infty} \int_{ \frac{\gamma}{ \rho } (1+\rho  x_A )}^{\infty} f(x_A,x_B) dx_A dx_B   \nonumber \\
    & = \psi_{1,2}(0)+ \psi_{2,1}(0)
\end{align}
where $\psi_{i,j}(t)$ is defined as follows for $(i,j) \in  \{1,2\}^2$:
\begin{align} \label{Definition_h1}
\psi_{i,j}(t)  = \frac{\lambda_i e^{-\frac{\lambda_j \gamma}{\rho}}}{\lambda_i + \lambda_j \gamma} e^{-(\lambda_i + \lambda_j \gamma) t}.
\end{align}

For the weak user $A$, $\phi_{A,N}(\rho) $ is derived from $\phi_{B,N}(\rho) $ by adding the constraint $x_A \geq \frac{\gamma}{ \rho}$. Its analytical expression is as follows:
\begin{align}\label{Phi1NCase1}
\phi_{A,N}(\rho)& = \int_{\frac{\gamma}{ \rho}}^{\infty} \int_{ \frac{\gamma}{ \rho } (1+\rho  x_A )}^{\infty}  f(x_A,x_B) dx_A dx_B  \nonumber \\
& =  \psi_{1,2}\left(\frac{\gamma}{ \rho}\right) + \psi_{2,1}\left(\frac{\gamma}{ \rho}\right) .
\end{align}

 \subsubsection{OMA probabilities}
The probability not to be in outage for the strong user $B$ is: 
 \begin{align} \label{CalculPhi20}
    \phi_{B,O}(\rho)  &= \int_{\frac{\tilde{\gamma}}{2  \rho}}^{\infty}    f_{X_B}(x_B) dx_B \nonumber \\
    &=   -e^{-(\lambda_1+ \lambda_2)\frac{\tilde{\gamma}}{2  \rho}} + e^{-\lambda_1 \frac{ \tilde{\gamma}}{2 \rho}} +  e^{-\lambda_2 \frac{ \tilde{\gamma}}{2\rho}}.
\end{align}

And for the weak user $A$, it is equal to: 
  \begin{align} \label{CalculPhi10}
    \phi_{A,O}(\rho) =  \int_{\frac{ \tilde{\gamma}}{2  \rho}}^{\infty}  f_{X_A}(x_A) dx_A =  e^{-(\lambda_1+\lambda_2)\frac{\tilde{\gamma}}{ 2 \rho}}.
\end{align}

\subsection{Comparison of OMA and NOMA throughputs}

\begin{proposition} \label{PropUserA_Throughput}
Whatever the values of $\lambda_1, \lambda_2$ and $\gamma$, there exists a value $\rho_{\text{min}}$ such that  the OMA throughput is larger than the NOMA throughput for  user $A$  for any $\rho$ exceeding $\rho_{\text{min}}$. OMA is therefore more efficient  than NOMA at large average SNR for the weak user, whereas the opposite conclusion stands at low average SNR. 
\end{proposition}
\textit{\textbf{Proof}}: The proof is given in Appendix \ref{AppendixA}. 

\begin{remark} 
In NOMA, the outage probability of user $B$ is always lower than the outage probability of user $A$. Moreover, the outage probability of user $B$ is lower than that of user $A$ in OMA. 
\end{remark}
\begin{remark} \label{PropUserB_Throughput}
 The OMA throughput is larger than the NOMA throughput for user $B$ when   $\rho >>1$. The opposite conclusion stands when $\rho \rightarrow 0$.  This result combined with Prop. \ref{PropUserA_Throughput} proves that the sum throughput is larger with OMA than with NOMA when  $\rho >>1$, and lower with OMA than with NOMA when $\rho \rightarrow 0$.
\end{remark}

\subsection{Adaptive NOMA strategy \label{NOMAA-NoCSIT-Section}}
The analytical expressions of $\phi_A$ and $\phi_B$ allow us to define an MA strategy called NOMA-Adaptive (NOMA-A). It adaptively selects between OMA and NOMA with the objective to  maximize the sum throughput. NOMA-A selection only depends  on   the following large-scale parameters: $\left(\rho, P_1, P_2, \gamma\right)$. All of these parameters are known at the receiver and do not vary frequently, as only $P_1$ and $P_2$  may be updated  if users move and their path loss and shadowing change. $\rho$ may also be varying if we assume that some additional interference is included in the noise power. Then a centralized controller decides, depending on these parameters, if NOMA or OMA should be used, and sends a one-bit feedback to both users so that they should adapt their transmission strategy. If we assume that the  one-bit feedback is only sent whenever the MA strategy should be modified,  then the required feedback amount is very low.

\section{Performance analysis and adaptive NOMA strategy in the Full CSIT case \label{SectionFullCSIT} }

We now assume that both transmitters have full CSIT. Moreover, users have minimum instantaneous rates requirements, and they do not transmit if  their instantaneous capacity is below a minimum threshold equal to $R_{k,\text{min}} =\log_2(1+\gamma)$ for $k\in \{A,B\}$. As in the previous case, we set the constraint  $\gamma \geq 1$. In NOMA, this threshold is applied on the instantaneous capacity per time slot, corresponding to a minimum SINR per time slot, $\gamma$. 
However in OMA, as transmission occurs only once every two time slots per user, the instantaneous capacity  during the active time slot should be equal to $2 R_{k,\text{min}}$. Consequently, as in the no CSIT case, the minimum SNR in OMA is written as $\tilde{\gamma} = 2 \gamma + \gamma^2$ so that $\log_2\left(1+ \tilde{\gamma} \right)  = 2 R_{k,\text{min}}$. This is mandatory in order to have a fair comparison between NOMA and OMA. To summarize, users either transmit at their instantaneous capacity using  (\ref{R1NOMADefinition}), (\ref{R2NOMADefinition}) or (\ref{ROMA}) if their SINR exceeds $\gamma$ (in NOMA) or $\tilde{\gamma}$ (in OMA), or they are inactive.

\subsection{Closed-form expressions of the average data rate in OMA and NOMA}
In the following, we use the auxiliary function:
\begin{align} \label{UsefulIntegralGamma}
 \alpha(\gamma, \lambda , \rho) &= \int_{x=\frac{\gamma}{\rho}}^{\infty} \log\left(1 + \rho x\right) e^{-\lambda x} dx  \nonumber \\ &=  \frac{e^{-\lambda \frac{\gamma}{\rho}}}{\lambda} \log\left(1+\gamma\right) + \frac{e^{ \frac{\lambda}{\rho}}}{\lambda}  E_1\left(\frac{(\gamma+1)\lambda}{\rho}  \right) 
\end{align}
where $E_1(x) = \int_{x}^{\infty} \frac{e^{-t}}{t} dt$ is the exponential integral.

\subsubsection{NOMA average data rates}
In NOMA, the instantaneous capacity of user $B$ depend on that of user $A$: if $x_A \geq \frac{\gamma}{\rho}$ then user $A$ is active and user $B$ suffers from its interference. Therefore user $B$ is only active if $x_B \geq \frac{\gamma}{ \rho}(1+\rho x_A)$. However if  $x_A < \frac{\gamma}{\rho}$, then user $A$ is not active and the constraint becomes $x_B \geq  \frac{\gamma}{\rho}$ for user $B$ to be active. 

The average data rate of user $A$ in NOMA is:
\begin{align} \label{AverageDataRateUserNOMA1}
    \mathbb{E}[R_A]&  = \int_{\frac{\gamma}{\rho}}^{\infty} \log_2\left(1 + \rho x_A\right) f_{X_A}(x_A)   dx_A  \nonumber \\
    & =  \frac{1}{\log(2)} (\lambda_1+\lambda_2)  \alpha(\gamma, \lambda_1+\lambda_2 ,  \rho). 
\end{align}

The average data rate of user $B$ in NOMA is:
\begin{align} \label{AverageDataRateUserNOMA2}
   & \mathbb{E}[R_B]  = \int_{x_A=0}^{\frac{\gamma}{\rho}} \int_{\frac{\gamma}{ \rho}}^{\infty} \log_2\left(1 +  \rho x_B\right) f(x_A,x_B) dx_B   dx_A  \nonumber \\
    & + \int_{\frac{\gamma}{\rho}}^{\infty} \int_{\frac{\gamma}{ \rho}(1+\rho x_A)}^{\infty} \log_2\left(1 +  \frac{\rho x_B}{1+\rho x_A}\right) f(x_A,x_B) dx_B   dx_A \nonumber \\
    &  =J_{B/\bar{A}} + J_{B/A}.
\end{align}

$J_{B/\bar{A}}$ is the average data rate of user $B$ when user $A$ is inactive. Its closed-form expression is given by:  
\begin{align} \label{AverageDataRateUserNOMA2_Ia}
   J_{B/\bar{A}} &=   \int_{0}^{\frac{\gamma}{\rho}} \int_{\frac{\gamma}{ \rho}}^{\infty} \log_2\left(1 +  \rho x_B\right) f(x_A,x_B) dx_B   dx_A  \nonumber \\
   & =\frac{1}{\log(2)}\left(\lambda_2\left(1- e^{-\frac{\lambda_1 \gamma}{\rho}}\right) \alpha\left(\gamma,\lambda_2,\rho \right) \right. \nonumber \\
  & \left.+ \lambda_1\left(1- e^{-\frac{\lambda_2 \gamma}{\rho}} \right) \alpha\left(\gamma,\lambda_1,\rho \right) \right).
\end{align}

$J_{B/A}$ is the average data rate of user $B$ when user $A$ is active. It is written as:
\begin{align} \label{AverageDataRateUserNOMA2_Ib1erepartie}
 & J_{B/A} \nonumber  \\
 &=\int_{\frac{\gamma}{\rho}}^{\infty} \int_{\frac{\gamma}{ \rho}(1+\rho x_A)}^{\infty} \log_2\left(1 + \rho x_A + \rho x_B\right)  f(x_A,x_B) dx_B   dx_A \nonumber \\
   & \hspace{0.3cm} - \int_{\frac{\gamma}{\rho}}^{\infty} \int_{\frac{\gamma}{ \rho}(1+\rho x_A)}^{\infty} \log_2\left(1 + \rho x_A \right) f(x_A,x_B) dx_B   dx_A.
\end{align}

 Determining the closed-form expression of the first integral in (\ref{AverageDataRateUserNOMA2_Ib1erepartie})  requires to express the following integral of the exponential integral \cite{TableIntegralsEI69}: 
      \begin{align}  \label{LaplaceTransformShiftedE1}
      & \int_{x=S}^{\infty} e^{-px}\int_{y = ax+b}^{\infty} \frac{e^{-y}}{y}  dy dx \nonumber \\ & =\frac{e^{-pS}}{p} E_1(b+aS)-  \frac{e^{\frac{bp}{a}}}{p}E_1\left((b+aS)(1+\frac{p}{a}) \right).
    \end{align}
   We can notice that  (\ref{LaplaceTransformShiftedE1}) provides the Laplace transform of $E_1(ax+b)$ if parameters $(p, a, b)$ are positive and $S=0$. 
    
By inserting (\ref{LaplaceTransformShiftedE1}) in  (\ref{AverageDataRateUserNOMA2_Ib1erepartie}), we finally obtain the closed-form expression of  $J_{B/A}$:
    \begin{align} \label{AverageDataRateUserNOMA2_Ib1erepartieFinal}
 J_{B/A} & =    \frac{1}{\log(2)}\left(\beta_{1,2} + \beta_{2,1}  \right)
 \end{align}
 where  $\beta_{i,j}$ is defined as follows for $(i,j) \in  \{1,2\}^2$:
   \begin{align} \label{betadefinition}
 \beta_{i,j} & = \frac{\lambda_i e^{-\frac{\lambda_j \gamma}{\rho}}}{\lambda_i+\lambda_j \gamma}\log(1+\gamma) e^{-(\lambda_i+\lambda_j \gamma)\frac{\gamma}{\rho}}  \nonumber \\
 & + \frac{\lambda_j e^{\frac{\lambda_i}{\rho}}}{(\lambda_j - \lambda_i)}\left( e^{-(\lambda_j - \lambda_i)\frac{\gamma}{\rho}} E_1\left(\frac{\lambda_i}{\rho}(\gamma+1) (1+\gamma)\right)  \right. \nonumber \\
 & \left.- e^{\frac{(\lambda_j-\lambda_i)}{\rho}}E_1\left(\frac{(\lambda_i\gamma+\lambda_j)(1+\gamma)}{\rho} \right) \right). 
 \end{align}

 $\mathbb{E}[R_B]$ is then deduced from  (\ref{AverageDataRateUserNOMA2}), (\ref{AverageDataRateUserNOMA2_Ia}) and (\ref{AverageDataRateUserNOMA2_Ib1erepartieFinal}).

\subsubsection{OMA average data rates}
In OMA, the instantaneous capacities of users $A$ and $B$ are independent. For user $A$, the average data rate is then equal to:
\begin{align} \label{AverageDataRateUserOMA1}
    \mathbb{E}[\tilde{R}_A]&  = \int_{\frac{\tilde{\gamma}}{2 \rho}}^{\infty} \frac{1}{2}\log_2\left(1 + 2 \rho x_A\right) f_{X_A}(x_A) dx_A  \nonumber \\
    & =  \frac{1}{2 \log(2)} ( \lambda_1+\lambda_2) \alpha(\tilde{\gamma}, \lambda_1+\lambda_2 , 2 \rho). 
\end{align}
And the average data rate of user $B$ is:
\begin{align} \label{AverageDataRateUserOMA2}
    \mathbb{E}[\tilde{R}_B]&  = \int_{\frac{\tilde{\gamma}}{2 \rho}}^{\infty} \frac{1}{2}\log_2\left(1 + 2 \rho x_B\right) f_{X_B}(x_B) dx_B  \nonumber \\
    & = \frac{1}{2 \log(2)}  \lambda_1 \alpha(\tilde{\gamma}, \lambda_1 , 2 \rho)  + \frac{1}{2 \log(2)}  \lambda_2 \alpha(\tilde{\gamma}, \lambda_2 , 2 \rho)  \nonumber \\&  -  \frac{1}{2 \log(2)} ( \lambda_1+\lambda_2) \alpha(\tilde{\gamma}, \lambda_1+\lambda_2 , 2 \rho). 
\end{align}

\subsection{Adaptive NOMA strategy \label{NOMAA-FullCSIT-Section}}
Before comparing the analytical expressions of the average data rates with OMA and NOMA, we propose an adaptive NOMA strategy for the full CSIT case, that similarly to the algorithm detailed for the no CSIT case in section \ref{NOMAA-NoCSIT-Section}, adaptively chooses an MA strategy between NOMA and OMA depending on the system parameters. 

\subsubsection{Description of NOMA-A full CSIT algorithm}
The full CSIT NOMA-A strategy  assumes that a centralized controller has access to all CSI and takes a decision to transmit or not for both users and with which MA strategy. It selects the MA strategy that maximizes the probability that both users are simultaneously active.  In details, the algorithm proceeds as follows: if NOMA can be used with both users active, then NOMA is selected. If the strong user can be active but the weak user cannot, then the strong user transmits interference-free. If the strong user cannot be active in NOMA and the weak user is active, then the system chooses OMA strategy  if it allows  both users to be active. Otherwise, only one of the users is active. 

The NOMA-A algorithm  is written in pseudo-code in Algorithm  \ref{NOMA-A-Alg-Description}. 
     

\begin{algorithm}[htbp]
\SetAlgoLined

\label{NOMA-A-Alg-Description}
 \If{  $x_A \geq \frac{\gamma}{\rho}$ \text{ and }  $x_B \geq \frac{\gamma}{ \rho}(1+\rho x_A)$} {both users are active in NOMA\;}
 \If{  $x_A < \frac{\gamma}{\rho}$ \text{ and }   $x_B \geq \frac{\gamma}{ \rho}$} {user $A$ is inactive and user $B$ transmits interference-free\;}
 \If{$x_A \geq \frac{\gamma}{\rho}$ \text{ and }  $x_B < \frac{\gamma}{ \rho}(1+\rho x_A)$}  {check if OMA can be used\: \\
\If{  $x_A \geq \frac{\tilde{\gamma}}{2\rho}$ }
{
\If{$x_B \geq \frac{\tilde{\gamma}}{2\rho}$}  
{
both users are active in OMA\;
}
 \Else{ only user $A$ transmits interference-free;\ }
 }
 
 \Else{\If{$x_B \geq \frac{\gamma}{\rho}$}  {only user $B$ transmits interference-free\; 
 }
 \Else{ both users are inactive; }
}
}
\caption{NOMA-A Full CSIT algorithm}
\end{algorithm}

\subsubsection{Closed-form expressions of NOMA-A data rates}
We hereunder derive the closed-form expressions of the NOMA-A data rates in order to compare them with OMA and NOMA rates. \\
In NOMA-A, user $A$ is either active together with user $B$ in NOMA or active alone in OMA when the SINR of user $B$ is lower than $\gamma$. The average data rate of the weak user consequently is: 
\begin{align} \label{AverageDataRateUserNOMAA1}
   & \mathbb{E}[\hat{R}_A] \nonumber \\
   &= \int_{\frac{\gamma}{\rho}}^{\infty} \int_{\frac{\gamma}{ \rho}(1+\rho x_A)}^{\infty} \log_2\left(1 +  \rho x_A\right) f(x_A,x_B) dx_B   dx_A  \nonumber \\
    & + \int_{\frac{\tilde{\gamma}}{2\rho}}^{\infty} \int_{x_A}^{\frac{\gamma}{ \rho}(1+\rho x_A)} \frac{\log_2\left(1 + 2  \rho x_A\right)}{2} f(x_A,x_B) dx_B   dx_A 
    \nonumber \\
  &=   \frac{1}{\log(2)} \left( \chi_{1,2} + \chi_{2,1} \right)
    \end{align}
where  $\chi_{i,j}$ for $(i,j) \in  \{1,2\}^2$ is equal to:
\begin{align} \label{Definitionkchi}
    \chi_{i,j} &  =\lambda_i e^{-\lambda_j \frac{\gamma}{\rho}} \alpha(\gamma, \lambda_i+\lambda_j \gamma,  \rho)  \nonumber \\
    & +  \frac{\lambda_i}{2}\left(\alpha(\tilde{\gamma}, \lambda_i+\lambda_j,  2\rho)  - e^{-\lambda_j \frac{\gamma}{\rho}}  \alpha(\tilde{\gamma}, \lambda_i+\lambda_j \gamma,  2\rho)\right).
\end{align}

User $B$ is active in several situations: either in NOMA of OMA when user $A$ is  active as well, or when user $A$ is inactive. The strong user's average data rate is thus equal to:
\begin{align} \label{AverageDataRateUserNOMAA2}
   & \mathbb{E}[\hat{R}_B]  = \int_{0}^{\frac{\gamma}{\rho}} \int_{\frac{\gamma}{ \rho}}^{\infty} \log_2\left(1 +  \rho x_B\right) f(x_A,x_B) dx_B   dx_A  \nonumber \\
    & + \int_{\frac{\gamma}{\rho}}^{\infty} \int_{\frac{\gamma}{ \rho}(1+\rho x_A)}^{\infty} \log_2\left(1 +  \frac{\rho x_B}{1+\rho x_A}\right) f(x_A,x_B) dx_B   dx_A \nonumber \\
  &+  \int_{\frac{\gamma}{\rho}}^{\frac{\tilde{\gamma}}{2\rho}} \int_{x_A}^{\frac{\gamma}{ \rho}(1+\rho x_A)} \log_2\left(1 +  \rho x_B\right) f(x_A,x_B) dx_B   dx_A  \nonumber \\
    &  + \int_{\frac{\tilde{\gamma}}{2\rho}}^{\infty} \int_{x_A}^{\frac{\gamma}{ \rho}(1+\rho x_A)} \frac{1}{2}\log_2\left(1 + 2  \rho x_B\right) f(x_A,x_B) dx_B   dx_A \nonumber \\
& = \mathbb{E}[R_B] + \hat{J}_{B/\bar{A}}  +   \hat{J}_{B/A}. 
\end{align}
The average data rate of user $B$ in NOMA-A when using OMA and $A$ is active is denoted as $\hat{J}_{B/A}$. Its closed-form expression is as follows:
\begin{align}\label{AverageDataRateUserNOMAA2PartOMA}
\hat{J}_{B/A} = &\int_{\frac{\tilde{\gamma}}{2\rho}}^{\infty} \int_{x_A}^{\frac{\gamma}{ \rho}(1+\rho x_A)} \frac{\log_2\left(1 + 2  \rho x_B\right)}{2} f(x_A,x_B) dx_B   dx_A \nonumber \\
 & = \frac{1}{2\log(2)} \left(\omega_{1,2} + \omega_{2,1}\right)
 \end{align}
where  $\omega_{i,j}$ for $(i,j) \in  \{1,2\}^2$ is defined as:
 \begin{align}\label{omegeDefinition }
 & \omega_{i,j} = \lambda_i \alpha\left(\tilde{\gamma}, \lambda_i+\lambda_j, 2 \rho \right) \nonumber \\
 & -\frac{\lambda_i e^{-\frac{\lambda_j \gamma}{\rho}}}{(\lambda_i+\lambda_j \gamma)}  \left(  e^{-(\lambda_i+\lambda_j \gamma)\frac{\tilde{\gamma}}{2\rho}} \log\left(1+2\gamma+\gamma \tilde{\gamma} \right) \right) \nonumber\\
 &  -\frac{\lambda_i e^{-\frac{\lambda_j \gamma}{\rho}}}{(\lambda_i+\lambda_j \gamma)}  \left(  e^{(\lambda_i+\lambda_j \gamma)(\frac{1+2 \gamma}{2\rho \gamma})}  \right. \nonumber \\
 & \left. E_1\left((\lambda_i+\lambda_j \gamma)\left(\frac{\tilde{\gamma}}{2\rho}+\frac{1+2 \gamma}{2\rho \gamma} \right) \right)\right)\nonumber \\
 & + e^{-\lambda_i  \frac{\tilde{\gamma}}{2\rho}} E_1\left(\frac{\lambda_j}{2 \rho}(1+\tilde{\gamma}) \right) - e^{\frac{\lambda_i}{2 \rho}}E_1\left(\frac{(\lambda_i+\lambda_j)}{2 \rho} (1+\tilde{\gamma}) \right) \nonumber\\
 &- e^{-\lambda_i  \frac{\tilde{\gamma}}{2\rho}} E_1\left(\frac{\lambda_j \gamma}{ \rho}\left(1+\frac{\tilde{\gamma}}{2}+\frac{1}{2\gamma}\right) \right)\nonumber \\
 &+  e^{\frac{\lambda_i}{\rho} \left(1 +\frac{1}{2 \gamma} \right)}E_1\left(\frac{(\lambda_j \gamma +\lambda_i)}{ \rho}\left(1+\frac{\tilde{\gamma}}{2}+\frac{1}{2\gamma}\right) \right).
 \end{align}

Finally, the average data rate of user $B$ in NOMA-A when using OMA and $A$ is inactive, denoted as  $\hat{J}_{B/\bar{A}}$,  is equal to:
 \begin{align}  \label{AverageDataRateUserNOMAA2PartOMAAInactive}
& \hat{J}_{B/\bar{A}}   \nonumber \\
&= \frac{1}{\log(2)}
 \int_{\frac{\gamma}{\rho}}^{\frac{\tilde{\gamma}}{2\rho}} \int_{x_A}^{\frac{\gamma}{ \rho}(1+\rho x_A)} \log_2\left(1 +  \rho x_B\right) f(x_A,x_B) dx_B   dx_A   \nonumber \\
 & = \frac{1}{\log(2)}\left(\delta(\gamma) - \delta\left(\frac{\tilde{\gamma}}{2}\right)   + \mu_{1,2}(\delta(\gamma) - \mu_{1,2}\left(\frac{\tilde{\gamma}}{2}\right) \right. \nonumber \\
 & \left. + \mu_{2,1}(\delta(\gamma) - \mu_{2,1}\left(\frac{\tilde{\gamma}}{2}\right) \right) 
 \end{align}

 where:
 \begin{align} \label{DefinitionDelta}
     \delta(t) & = e^{-(\lambda_1+\lambda_2) \frac{t}{\rho}} \log\left(1+t \right) \nonumber\\
& + e^{\frac{(\lambda_1+\lambda_2)}{\rho}} E_1\left(\frac{(\lambda_1+\lambda_2)}{\rho}(t+1) \right) 
 \end{align}
 and  $\mu_{i,j}$ for $(i,j) \in  \{1,2\}^2$ is defined as:
 \begin{align}\label{Definitionmu}
   &  \mu_{i,j}(t) =  \frac{\lambda_i e^{-\frac{\lambda_j\gamma}{\rho}}}{\lambda_i+\lambda_j\gamma} e^{-(\lambda_i+\lambda_j\gamma)\frac{t}{\rho}}  \log\left(1+\gamma+\gamma t \right)\nonumber \\
 & - \frac{\lambda_i e^{-\frac{\lambda_j\gamma}{\rho}}}{\lambda_i+\lambda_j\gamma} e^{(\lambda_i+\lambda_j\gamma)\frac{(1+\gamma)}{\gamma \rho }}  \nonumber \\
 & E_1\left(\frac{(\lambda_i+\lambda_j\gamma)}{\rho} \left(t+\frac{(1+\gamma)}{\gamma} \right)\right) \nonumber \\
  & + e^{-\lambda_i \frac{t}{\rho}} E_1\left( \frac{\lambda_j}{\rho}(1+t) \right) - e^{\frac{\lambda_1}{\rho}} E_1\left( \frac{(\lambda_i+\lambda_j)}{\rho}(1+t) \right) \nonumber \\
  & - e^{-\lambda_i \frac{t}{\rho}} E_1\left( \frac{\lambda_j}{\rho}(1+\gamma(1+t)) \right) \nonumber \\
 &  + e^{\frac{\lambda_i}{\rho}\left(1+\frac{1}{\gamma} \right)} E_1\left( \frac{\lambda_j}{\rho}(1+\gamma(1+t))(1+\frac{\lambda_i}{\lambda_j\gamma}) \right).
 \end{align}

 \subsection{Analytical comparison of average data rates for any $\rho$ regime}
  \begin{theorem}\label{PropDataRateWeakUserAnyRegimeFullCSIT}
The average data rate of user $A$ is always larger with NOMA than with NOMA-A and is always larger with NOMA-A than with OMA.
  \begin{align}\label{DataRateOrderingWeakUser}
    \mathbb{E}[R_A] \geq  \mathbb{E}[\hat{R}_A]  \geq \mathbb{E}[\tilde{R}_A]
\end{align}
   \end{theorem}
 \textit{\textbf{Proof}}: The proof is detailed in Appendix \ref{AppendixC}.
 
 \begin{theorem}\label{PropDataRateStrongUserAnyRegimeFullCSIT}
 The average data rate of  user $B$ in NOMA-A is always larger than with NOMA.
 \end{theorem}
  \textit{\textbf{Proof}}:   (\ref{AverageDataRateUserNOMAA2}) shows that $\mathbb{E}[\hat{R}_B]$ is the summation of $\mathbb{E}[R_B]$ and other positive terms. Therefore:
    \begin{align}\label{DataRateOrderingStrongUser}
    \mathbb{E}[\hat{R}_B] \geq \mathbb{E}[R_B].
\end{align}

 \subsection{Asymptotic behavior of the average data rates in $\rho$}
 \subsubsection{Asymptotic behavior in $\rho$ for user A}
 The exponential integral function is approximated as follows \cite{HandbookMathematicalFunctions}:
\begin{align} \label{E1Bounds}
  E_1(x) \approx  -\log(x) - \gamma_{E} - \sum_{n=1}^{\infty}(-1)^n\frac{x^n}{n n!}
\end{align}
where $\gamma_{E} \approx 0.57721$ is Euler's constant. 
When $\rho  >> 1$, $\sum_{n=1}^{\infty}\frac{(-1)^n }{{\rho}^n n n!} \rightarrow 0$. Consequently: 
 \begin{align} \label{E1Approx}
     E_1\left(\frac{a}{b \rho}\right) \approx \log(\rho) -\log\left(\frac{a}{b}\right) - \gamma_{E}.
 \end{align}
 Numerical observations show that this approximation is tight at large values of $\rho$. 
 The definition of function $\alpha$ (see  (\ref{UsefulIntegralGamma})) then involves that $\alpha$ can be approximated  when $\rho >> 1$ by:
\begin{align} \label{AlphaAsymptoticBehavior}
    \alpha(\gamma, \lambda,  \rho) \approx \frac{1}{\lambda} \log(\rho) - \frac{1}{\lambda}\left(\log(\lambda) + \gamma_{E}\right).
\end{align}
Consequently, both $\alpha$ and $E_1$ functions  have a linear behavior in $\log(\rho)$ at large average SNR. As all closed-form expressions of the average data rates  at large average SNR are linear combinations $\alpha$ and $E_1$, the average data rates are asymptotically linear in $\log(\rho)$ as well. 

\begin{lemma} \label{PropAsymptoticDataRateUserA}
The average data rate of user $A$ is a linear function of $\log(\rho)$ when $\rho >>1$.  Then the slope of the average data rate of user $A$ in $\log(\rho)$ is larger with NOMA than with NOMA-A, and larger with NOMA-A than with OMA. The slope in $\log(\rho)$ of the NOMA  average data rate of user $A$  is twice that of the OMA data rate ($\frac{1}{\log(2)}$ vs. $\frac{1}{2\log(2)}$), and the slope of the NOMA-A data rate is given by \begin{align}\frac{1}{2\log(2)}\left(1+\frac{\lambda_1}{\lambda_1+\lambda_2 \gamma} + \frac{\lambda_2}{\lambda_2+\lambda_1 \gamma} \right).
\end{align}
\end{lemma}
  \textit{\textbf{Proof}}: The proof is given in Appendix \ref{AppendixD}. 

We can notice that these results are consistent with Theorem \ref{PropDataRateWeakUserAnyRegimeFullCSIT}.

  \subsubsection{Asymptotic behavior in $\rho$ for the strong user (user B)}
  \begin{lemma} \label{PropAsymptoticDataRateUserB}
  The average data rate of user $B$ is a linear function of $\log(\rho)$ when $\rho >>1$.  Then the slope of the average data rate of user $B$ in $\log(\rho)$ is larger with OMA than with NOMA-A and with NOMA. The asymptotic slope in $\log(\rho)$ of the average data rate of user $B$  is equal to $\frac{1}{2\log(2)}$ in OMA, to  $0$ in NOMA and to: 
  \begin{align}
  \frac{1}{2\log(2)}
\left(1 - \frac{\lambda_1}{\lambda_1 + \lambda_2 \gamma} - \frac{\lambda_2}{\lambda_2 + \lambda_1 \gamma}\right)
\end{align}
in NOMA-A. 
  \end{lemma}
  
  \begin{proposition} \label{PropAsymptoticDataRateUserBAsymptote}
  The average data rate of user $B$ in NOMA has an asymptote when $\rho >>1$ which is equal to:
    \begin{align} \label{IbAsymptoticBehavior}
    \mathbb{E}[R_B] & \approx \frac{1}{\log(2)} \log(1+\gamma)\left(\frac{\lambda_1}{\lambda_1+ \lambda_2 \gamma} +\frac{\lambda_2}{\lambda_2+ \lambda_1 \gamma}  \right)   \nonumber \\
      & +  \frac{1}{\log(2)}\frac{1}{(\lambda_1 - \lambda_2)} \left(\lambda_1\log\left(\frac{\lambda_1+ \lambda_2 \gamma}{\lambda_1(1+\gamma)} \right)  \right.  \nonumber \\
      & - \left. \lambda_2\log\left(\frac{\lambda_2+ \lambda_1 \gamma}{\lambda_2(1+\gamma)} \right) \right).
  \end{align}
  \end{proposition}
    \textit{\textbf{Proof}}: The proofs of Lemma \ref{PropAsymptoticDataRateUserB} and Proposition \ref{PropAsymptoticDataRateUserBAsymptote} are given in Appendix \ref{AppendixE}.
    
  \subsubsection{Asymptotic behavior in $\rho$ for the sum data rate \label{SectionAsymptoticSumRate}}
  \begin{proposition}  \label{PropAsymptoticSumDataRate}
    All three strategies  have the same asymptotic slope in  $\log(\rho)$ in terms of sum data rate. Their slope is in $\frac{1}{\log(2)}$. 
  \end{proposition}
      \textit{\textbf{Proof}}: 

The asymptotic sum data rate in  $\log(\rho)$ with OMA, NOMA and NOMA-A, respectively, are  equal to:
  \begin{align}
       \mathbb{E}[\tilde{R}_A]+ \mathbb{E}[\tilde{R}_B] \approx \frac{1}{\log(2)} \log(\rho) - f_{\text{O}}\left(P_1, P_2,\gamma \right),
  \end{align}
    \begin{align}
       \mathbb{E}[R_A]+ \mathbb{E}[R_B] \approx \frac{1}{\log(2)} \log(\rho) - f_{\text{N}}\left(P_1,P_2,\gamma \right),
  \end{align}
 \begin{align} \label{SumRateNOMAA-Asymptotic}
      \mathbb{E}[\hat{R}_A] +  \mathbb{E}[\hat{R}_B]  \approx   \frac{1}{\log(2)} \log(\rho) - f_{\text{A}}\left(P_1,P_2,\gamma \right)
  \end{align}
  where  $f_{\text{O}}$, $f_{\text{N}}$ and $f_{\text{A}}$ express  constant terms for  OMA, NOMA and NOMA-A, respectively. They are not written here in order to avoid redundancy but they can be  obtained from the expressions of the average data rates. They depend on $P_1, P_2$ and $\gamma$. The expressions of  $f_{\text{O}}$, $f_{\text{N}}$ and $f_{\text{A}}$ do not simplify enough to deduce analytical expressions of the values of $\left(P_1, P_2, \gamma \right)$  for which NOMA-A provides  the largest asymptotic sum data rate. They are  however numerically evaluated in Section \ref{SectionResultsFullCSIT}. 
  
 \section{Simulation results \label{SectionResults}}
In this section, unless otherwise stated, the power values are set to $P_1 = 0.1$ and $P_2=0.9$ and the SINR thresholds to $\gamma = 10$ dB. 

\subsection{No CSIT case}
\begin{figure}[!ht]
\centering
  \includegraphics[width=3.5in]{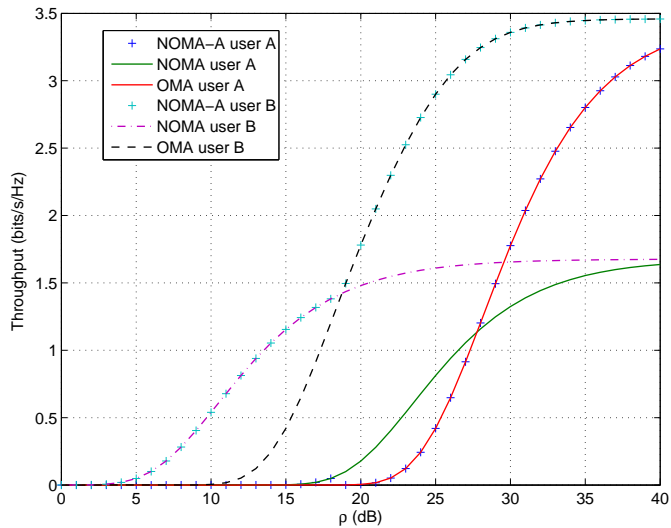}
  \vspace{-10pt}
  \caption{Throughput  per user vs $\rho$ when $\gamma = 10$ dB, $P_1 = 0.1$ and $P_2=0.9$ }\label{Throughput10dB0109}
\end{figure}

\begin{figure}[!ht]
\centering
  \includegraphics[width=3.5in]{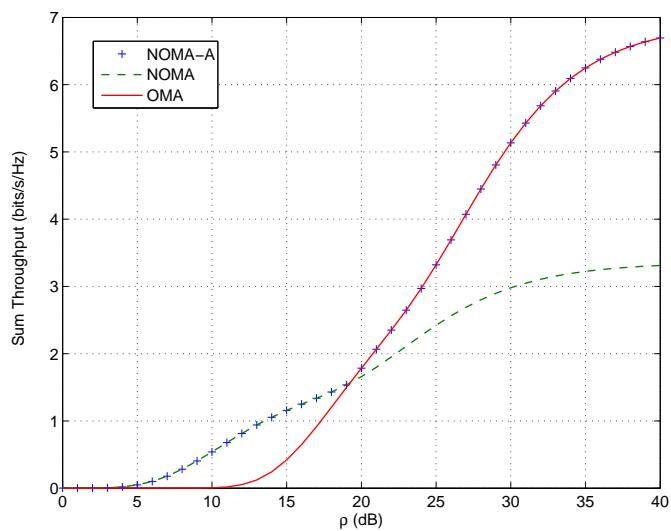}
  \vspace{-10pt}
  \caption{Sum throughput vs $\rho$ when $\gamma = 10$ dB, $P_1 = 0.1$ and $P_2=0.9$ }\label{SumThroughput10dB0109}
\end{figure}

\begin{figure}[!ht]
\centering
  \includegraphics[width=3.5in]{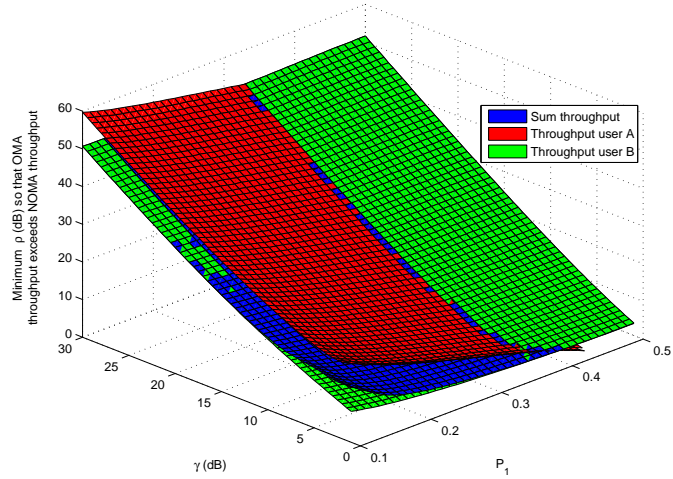}
  \vspace{-10pt}
  \caption{Minimum $\rho$ for the throughput to be larger with OMA than with OMA vs  $\gamma$  and $P_1$}\label{MinRho_3D}
\end{figure}

 Fig. \ref{Throughput10dB0109} and \ref{SumThroughput10dB0109} show the individual throughputs and the sum throughput depending on $\rho$, respectively, in the no CSIT case. As proven in  Proposition \ref{PropUserA_Throughput} and Remark \ref{PropUserB_Throughput}, OMA outperforms NOMA at large average SNR and NOMA outperforms OMA at low  average SNR. Moreover, NOMA-A is not only the best strategy in terms of sum throughput, but it is also observed that it leads to the largest throughput for the strong user. 
  Fig. \ref{MinRho_3D} represents the minimum value of $\rho$, denoted by $\rho_{\text{min}}$, such that OMA throughput is  larger than NOMA throughput for any $\rho \geq \rho_{\text{min}}$. We assume  that $P_2 = 1-P_1$. For low values of $P_1$,  $\rho_{\text{min}}$ is larger for user $A$ than for user $B$, whereas the opposite conclusion stands for larger values of $P_1$. On the one hand, when $P_1$ is low, user $A$ is mostly in outage due to its SNR constraint and not because of an outage of user $B$. The OMA SNR constraint $x_A \geq \frac{\tilde{\gamma}}{2 \rho}$ is indeed more stringent than the NOMA SNR constraint  $x_A \geq \frac{\gamma}{\rho}$ because $\tilde{\gamma} > 2 \gamma$ (see (\ref{EquivalentOMASNR})).   On the other hand,  when $P_1$ is large, user $B$  becomes interference-limited with NOMA and its outage also triggers an outage of user $A$. This also explains why the value of $\rho_{\text{min}}$ is almost similar for the sum throughput and for the throughput of user $B$ when $P_1$ is large. 
 
\subsection{Full CSIT case \label{SectionResultsFullCSIT}}
 
 \begin{figure}[!ht]
\centering
  \includegraphics[width=3.5in]{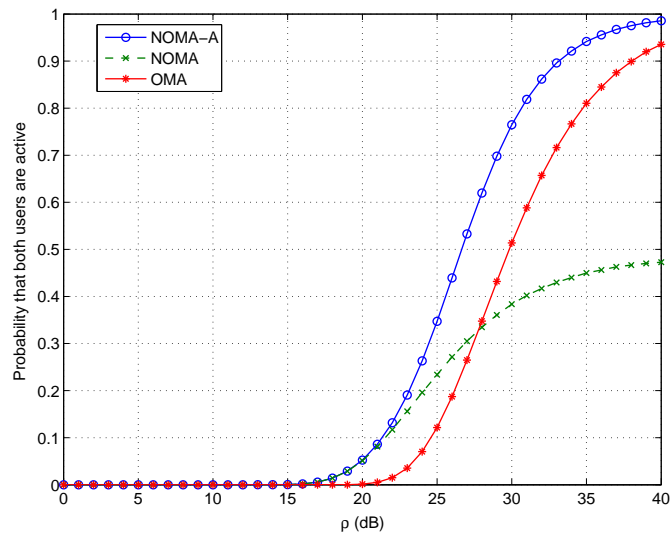}
  \vspace{-10pt}
  \caption{Probability that both users are active vs $\rho$ when $\gamma= 10$ dB, $P_1 = 0.1$ and $P_2=0.9$ }\label{ActivityProbability}
\end{figure}
   \begin{figure}[!ht]
\centering
  \includegraphics[width=3.5in]{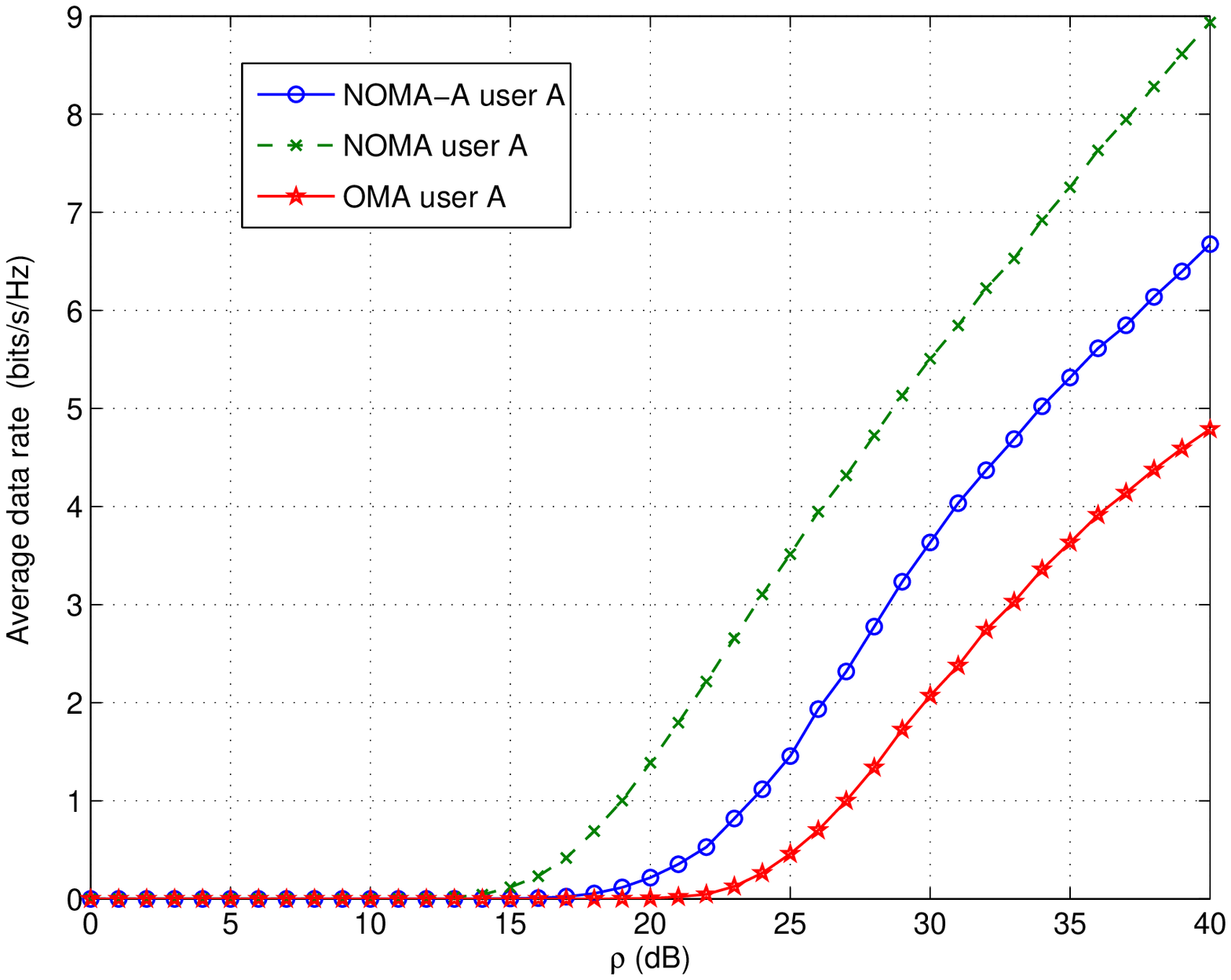}
  \vspace{-10pt}
  \caption{Data rate of the user $A$ vs $\rho$, $\gamma = 10$ dB, $P_1 = 0.1$ and $P_2=0.9$ }\label{RateNOMAAFullCSITUserA}
\end{figure}
 
  \begin{figure}[!ht]
\centering
  \includegraphics[width=3.5in]{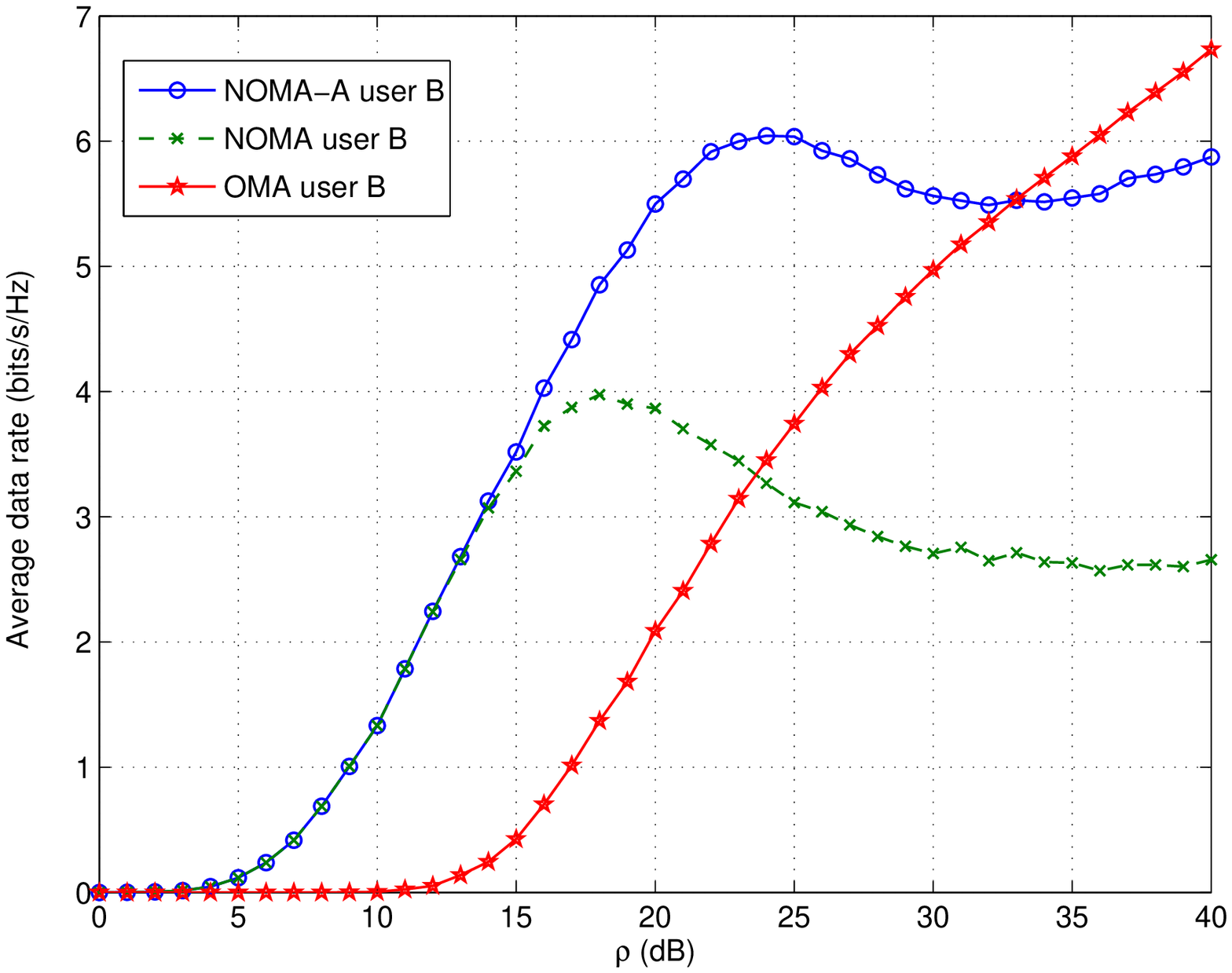}
  \vspace{-10pt}
  \caption{Data rate of the user $B$ vs $\rho$, $\gamma = 10$ dB, $P_1 = 0.1$ and $P_2=0.9$ }\label{RateNOMAAFullCSITUserB}
\end{figure}

 \begin{figure}[!ht]
\centering
  \includegraphics[width=3.5in]{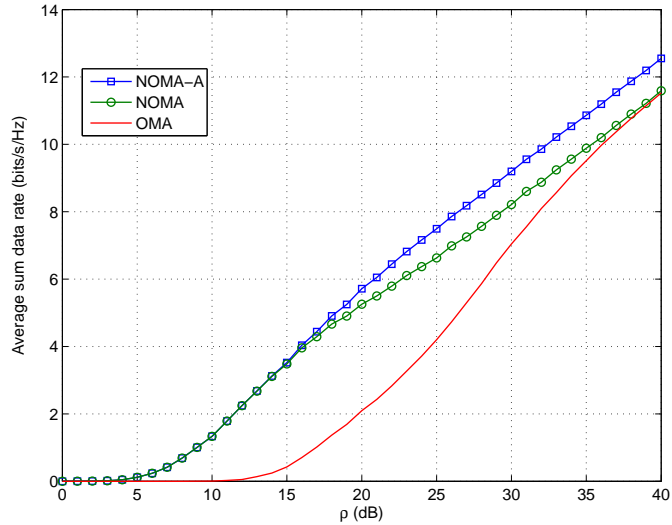}
  \vspace{-10pt}
  \caption{Sum data rate  vs $\rho$, $\gamma = 10$ dB, $P_1 = 0.1$ and $P_2=0.9$ }\label{Rates_NOMAAFullCSIT}
\end{figure}

 \begin{figure}[!ht]
\centering
  \includegraphics[width=3.5in]{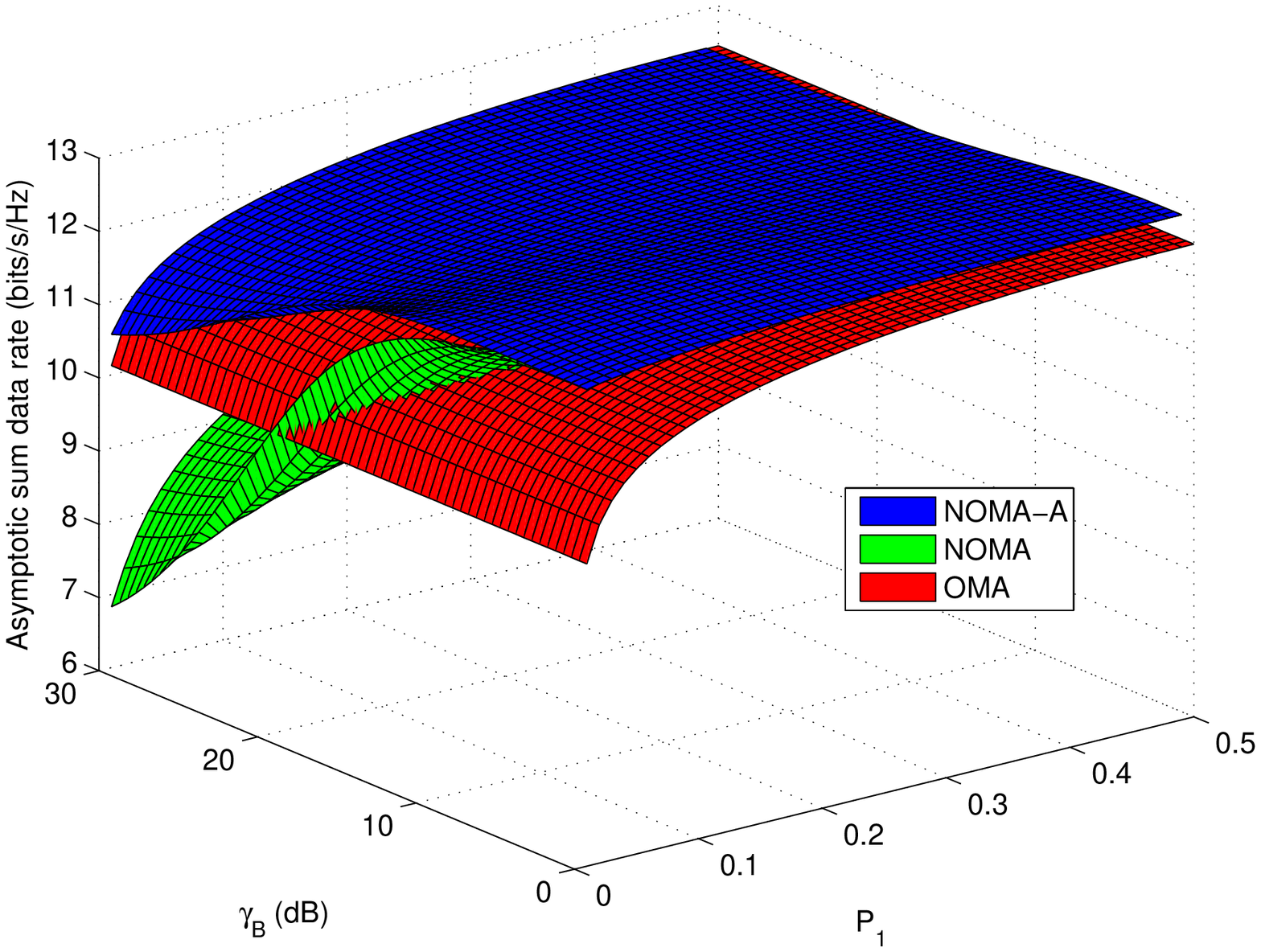}
  \vspace{-10pt}
  \caption{Asymptotic sum data rate  vs $\gamma$  and $P_1$ when $ \gamma = 10$ dB, $\rho = 40$ dB }\label{SumDataRateAsymptotic3D}
\end{figure}
In the full CSIT case, Fig. \ref{ActivityProbability} shows that NOMA-A maximizes the probability that both users are active, whatever the value of $\rho$.  Fig. \ref{RateNOMAAFullCSITUserA} confirms the theoretical results of Theorem \ref{PropDataRateWeakUserAnyRegimeFullCSIT}  and Lemma \ref{PropAsymptoticDataRateUserA}, showing that the  average data rate of the weak user is larger with NOMA than with NOMA-A, and larger with NOMA-A than with OMA, with the asymptotic slopes in $\log(\rho)$ given by Lemma \ref{PropAsymptoticDataRateUserA}. 
The results on Fig. \ref{RateNOMAAFullCSITUserB} are consistent with Theorem \ref{PropDataRateStrongUserAnyRegimeFullCSIT} and Lemma \ref{PropAsymptoticDataRateUserB}: the average data rate of the strong user is larger with NOMA-A than with NOMA, and the asymptotic slope of the average data rate in $\log(\rho)$ is larger  with OMA than with NOMA-A, whereas NOMA has an asymptote.  We can notice that NOMA-A outperforms OMA at low to medium values of $\rho$.
The sum data rate is larger with NOMA-A than with the two other strategies whatever the value of $\rho$, as shown by Fig. \ref{Rates_NOMAAFullCSIT}. Moreover, all  three strategies have the same asymptotic slope in  $\log(\rho)$,  as expected according to  Theorem \ref{PropAsymptoticSumDataRate}. 
Finally, Fig. \ref{SumDataRateAsymptotic3D} shows the asymptotic behavior of the sum data rate depending on $P_1$ and $\gamma$, assuming that $P_2 = 1 - P_1$. $\rho$ is set to $40$ dB and $\gamma$ to $10$ dB.  Fig. \ref{SumDataRateAsymptotic3D} shows that NOMA-A provides the largest sum data rate  whatever the parameter values, and that OMA outperforms NOMA when $\gamma$ or $P_1$ is large. A large  value of $P_1$ involves a large interference level on the strong user in NOMA, and the minimum instantaneous capacity becomes even more difficult to reach when  $\gamma$ increases, thus leading to many situations where user $B$ is not active. OMA avoids interference-limited situations, as well as NOMA-A since it then selects OMA instead of NOMA. These results further emphasize the relevance of adapting the MA strategy. 

\subsection{Extension to $K>2$ users \label{SectionResults3Users}}
In this section, we extend NOMA-A algorithms for $K>2$ ordered users. To the best of our knowledge, it is then not possible to obtain closed-form expressions of throughputs and average data rates when $K>2$. Even in the simplest case where all users have the same power, implying that $\{x_i\}_{1\leq i \leq K}$ are independent and identically distributed, closed-form expressions cannot be obtained with order statistics theory \cite{yang_alouini_2011}. This would indeed require expressing the joint pdf of one order statistics (which is not necessarily the largest) and of the  partial sum of lower order statistics. According to \cite{yang_alouini_2011}, these joint pdf are not known for all cases yet. Moreover, as explained in \cite{KoAlouiniMRC07}, even if closed-form expressions of these joint pdf were obtained, no closed-form expressions of the throughputs and average data rates could be directly deduced from them, and some mathematical tool would still be required to compute them. In the scenario considered in this paper, $\{x_i\}_{1\leq i \leq K}$ are not identically distributed because $\{P_i\}_{1\leq i \leq K}$ may take any value. This renders the problem even more difficult. For all these reasons, we rely to Monte Carlo simulations in this section to evaluate NOMA-A performance when $K>2$.

\begin{figure}[!ht]
\centering
  \includegraphics[width=3.5in]{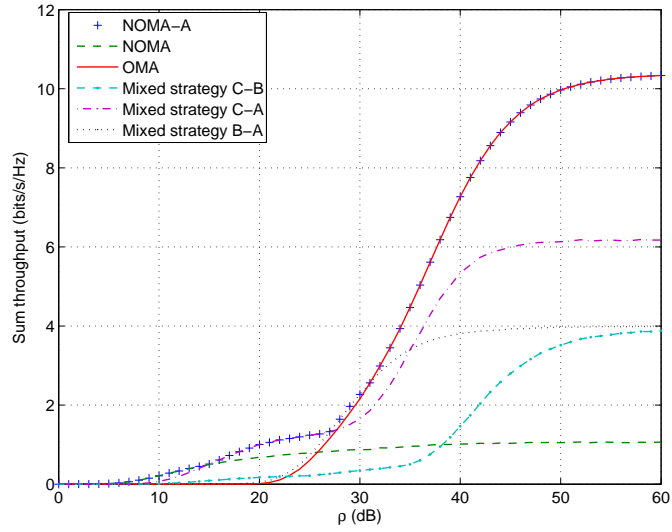}
  \vspace{-10pt}
  \caption{Sum throughput vs $\rho$, when $\gamma = 10$ dB, $K=3$, no CSIT }\label{SumThroughput3users}
\end{figure}

In the no CSIT case, NOMA-A algorithm can be extended as follows: the best MA strategy is the one that maximizes the sum throughput among OMA, NOMA, and among all the possible mixed strategies where a subset of users are involved in NOMA, while another subset is in OMA. For instance, if we consider $K=3$ with users $A, B, C$, assuming that $x_C \geq x_B \geq x_A$, three mixed strategies are possible. In each of them, two users  transmit in NOMA during $2$ time slots over $3$ using $3/2$ times their power per time slot  for fair comparison with NOMA, while the other user transmits in OMA during $1$ time slot over $3$, using $3$ times its power. Fig. \ref{SumThroughput3users} shows the sum throughput when $K=3$, where 'Mixed strategy C-B' in the legend means that users $C$ and $B$ are involved in NOMA.  The simulation parameters are $\gamma=10$ dB for all users, $P_1 = 0.05$, $P_2=0.15$ and $P_3 = 0.8$. Similarly to the two users case, NOMA is the best strategy at very low SNR and OMA is the best strategy at large SNR. However, at medium SNR, Mixed strategies involving users $C$ and $A$ in NOMA or involving users $B$ and $A$  provide larger sum throughputs.
 
\begin{figure}[!ht]
\centering
  \includegraphics[width=3.5in]{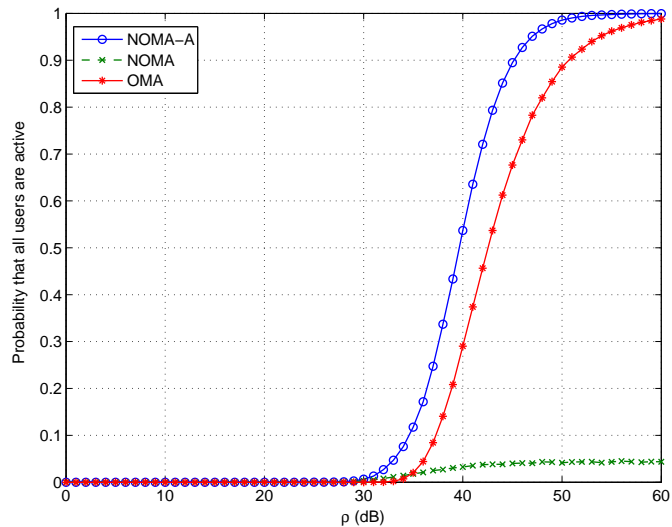}
  \vspace{-10pt}
  \caption{Probability that all users are   active vs $\rho$, when $\gamma = 10$ dB, $K=3$, full CSIT }\label{ActivityProbability3Users}
\end{figure}

In the full CSIT case, for each instantaneous channel realization, NOMA, OMA or one of the Mixed strategies is selected, with the objective to maximize the probability that all $K$ users are jointly active. Fig. \ref{ActivityProbability3Users} represents this probability when $K=3$, with the same parameters in the no CSIT case. We can notice that with NOMA, the probability to have all users active is very low, due to large interference. These numerical results assess that NOMA-A can be efficiently extended to larger number of users.

\section{Conclusions \label{SectionConclusions}}
This paper has focused on uplink two-user MA with outage situations in the no CSIT case and with a minimum instantaneous capacity threshold in the full CSIT case.  All analytical throughput and average data rate expressions in OMA and NOMA and in the proposed NOMA-A strategy have been obtained and compared. The superiority of NOMA-A in terms of sum throughput and sum data rate has been established. Future work will consist in considering other optimization objectives such as energy efficiency.


\appendices
\section{\label{AppendixA}}
By  (\ref{ThroughputOMA}) and (\ref{ThroughputNOMA}), OMA and NOMA throughputs can be directly compared by evaluating the sign of  $ \left(\frac{\phi_{A,N}(\rho)}{\phi_{A,O}(\rho) }  - 1\right)$.
Then: 
\begin{align}\label{RatioPhiA}
   g_A(\rho) &=  \frac{\phi_{A,N}(\rho)}{\phi_{A,O}(\rho) } \nonumber \\ 
   & = \frac{\lambda_1 }{\lambda_1 + \lambda_2 \gamma} e^{-\frac{(\lambda_2-\lambda_1) \gamma^2}{2\rho}} + \frac{\lambda_2 }{\lambda_2 + \lambda_1 \gamma} e^{-\frac{(\lambda_1-\lambda_2) \gamma^2}{2\rho}}.
\end{align}
The asymptotic behavior of $g_A(\rho)$ is as follows:  $\ g_A(\rho)>>1 $ when $\rho \rightarrow 0$ and   $\ g_A(\rho)\rightarrow m(\gamma)$ when $\rho >>1$, where
\begin{align} \label{definitionmgamma}
m(\gamma)= \frac{\lambda_1 }{\lambda_1 + \lambda_2 \gamma} + \frac{\lambda_2 }{\lambda_2 + \lambda_1 \gamma}.
\end{align}
 $m(\gamma)$ is a decreasing function of $\gamma$ and $m(1) = 1$. Consequently, as we assumed that $\gamma \geq 1$,   $m(\rho) \leq 1$.\\
We hereafter prove that $g_A(\rho)$ is a decreasing function in $\rho$. Its first derivative is equal to:
\begin{align}
    g_A'(\rho) & = \frac{\gamma^2}{\rho^2} e^{-\frac{(\lambda_2-\lambda_1) \gamma^2}{2\rho}} 
    \left(\frac{\lambda_1(\lambda_2-\lambda_1) }{\lambda_1 + \lambda_2 \gamma}  \right. \nonumber \\
    &\left. +  \frac{\lambda_2 (\lambda_1-\lambda_2)}{\lambda_2 + \lambda_1 \gamma} e^{-\frac{-(\lambda_2-\lambda_1) \gamma^2}{\rho}}  \right).
\end{align}
As  $e^{-\frac{(\lambda_2-\lambda_1) \gamma^2}{\rho}} \leq 1$,  $g_A'(\rho)$ is negative  if $\left(\frac{\lambda_1(\lambda_2-\lambda_1) }{\lambda_1 + \lambda_2 \gamma}   +  \frac{\lambda_2 (\lambda_1-\lambda_2)}{\lambda_2 + \lambda_1 \gamma}\right)$ is negative. The sign of $\left(\frac{\lambda_1(\lambda_2-\lambda_1) }{\lambda_1 + \lambda_2 \gamma}   +  \frac{\lambda_2 (\lambda_1-\lambda_2)}{\lambda_2 + \lambda_1 \gamma}\right)$ is equivalent to the sign of $\rho \lambda_1^2 (\lambda_2 - \lambda_1) \left(1-\left(\frac{\lambda_2}{\lambda_1}\right)^2 \right)$ which is always negative because $P_2 \geq P_1$, which implies that $\lambda_2 - \lambda_1 \leq 0$ and $1-\left(\frac{\lambda_2}{\lambda_1}\right)^2\geq 0$. 

To conclude, $g_A(\rho)$ is a decreasing function in $\rho$ that tends to a value which is lower than $1$. Consequently, whatever the values of $\lambda_1, \lambda_2$ and $\gamma$, there exists a value $\rho_{\text{min}}$ such that  $\phi_{A,O}(\rho) \geq  \phi_{A,N}(\rho)$ for any $\rho \geq \rho_{\text{min}}$. By the definition of the throughputs (\ref{ThroughputOMA}), (\ref{ThroughputNOMA}), this implies that the throughput  is larger  with OMA than with NOMA for any  $\rho \geq \rho_{\text{min}}$. This completes the proof. 

The proof of Remark 2 is given hereunder:\\
For the strong user $B$, the asymptotic behavior of $\phi_B$ is as follows: 
\begin{itemize}
    \item  $\phi_{B,N} \rightarrow m(\gamma) \leq 1$ when $\rho >>1$;
    \item $\phi_{B,O} \rightarrow 1$ when $\rho >>1$. Consequently, $\phi_{B,O} \geq \phi_{B,N}$ when $\rho >>1$;
    \item Both $\phi_{B,N}$ and $\phi_{B,O}$ tend to $0$ when $\rho \rightarrow 0$ but  the decrease of $\phi_{B,O}$ towards $0$ is faster than that of $\phi_{B,N}$ due to the extra term $-e^{-(\lambda_1+\lambda_2) \frac{\tilde{\gamma}}{2 \rho}}$ in $\phi_{B,O}$.
\end{itemize}
Therefore we can conclude that OMA is more efficient than NOMA for the strong user at large average SNR, whereas the opposite stands at low average SNR.

\section{ \label{AppendixC}}
By definition, the average data rate of the weak user with NOMA-A is:
\begin{align} \label{RateUserANOMAA_ConditionalExp}
    \mathbb{E}[\hat{R}_A]  & = \mathbb{E}\left[R_A  \ | \ (x_B \geq \frac{\gamma}{ \rho}(1+\rho x_A))\right]  \nonumber \\
    & + \mathbb{E}\left[\tilde{R}_A   \ | \  (x_B < \frac{\gamma}{ \rho}(1+\rho x_A))\right]. 
\end{align}
The instantaneous data rate of user $A$  is always larger with NOMA than with OMA as $\log_2\left(1+\rho x_A\right) \geq \frac{1}{2}\log_2\left(1+2 \rho x_A\right)$. Using $R_A \geq \tilde{R}_A$ and the monotonicity of conditional expectations, we have that:
\begin{align} \label{ComparisonNOMAandNOMAAUserA}
    \mathbb{E}[\hat{R}_A]  &\leq \mathbb{E}\left[R_A \ | \ (x_B \geq \frac{\gamma}{ \rho}(1+\rho x_A))\right]  \nonumber \\
    &+ \mathbb{E}\left[R_A   \ | \ (x_B < \frac{\gamma}{ \rho}(1+\rho x_A))\right]  =  \mathbb{E}[R_A ]
\end{align}.
And similarly:
\begin{align} \label{ComparisonOMAandNOMAAUserA}
    \mathbb{E}[\hat{R}_A]  & \geq \mathbb{E}\left[\tilde{R}_A \ | \ x_B \geq (\frac{\gamma}{ \rho}(1+\rho x_A)\right)]  \nonumber \\
    &+ \mathbb{E}\left[\tilde{R}_A   \ | \ x_B < (\frac{\gamma}{ \rho}(1+\rho x_A))\right]  = \mathbb{E}[\tilde{R}_A].
\end{align}

The average data rate ordering for the weak user is therefore given by  (\ref{DataRateOrderingWeakUser}).

\section{\label{AppendixD}}
From \ref{AverageDataRateUserOMA1}) and (\ref{AlphaAsymptoticBehavior}), the asymptotic behavior when ${\rho  >> 1}$ of the OMA data rate for the weak user is:
\begin{align} \label{OMA-A-AsymptoticBehavior}
    \mathbb{E}[\tilde{R}_A]\approx \frac{1}{2 \log(2)}  \left( \log(\rho) + \log(2) - \log(\lambda_1+ \lambda_2) - \gamma_{E}  \right).
\end{align}

Similarly from (\ref{AverageDataRateUserNOMA1}) and (\ref{AlphaAsymptoticBehavior}), the asymptotic behavior of the NOMA data rate for the weak user is:
\begin{align} \label{NOMA-A-AsymptoticBehavior}
    \mathbb{E}[R_A]\approx \frac{1}{\log(2)}\left( \log(\rho) - \log(\lambda_1+ \lambda_2) - D\right).
\end{align}
Consequently the slope of the asymptotic data rates in $\log(\rho)$ with NOMA is twice that obtained with OMA. 

Finally from  (\ref{AverageDataRateUserNOMAA1}), the asymptotic behavior of the NOMA-A data rate for the weak user is:
\begin{align} \label{NOMAA-A-AsymptoticBehavior}
   &\mathbb{E}[\hat{R}_A]   \nonumber \\
   &\approx \frac{1}{2\log(2)}\left(1+m(\gamma) \right)  \left(\log(\rho) - \gamma_{E}\right)  \nonumber \\
   &+ \frac{1}{2}\left(1-\frac{\lambda_1}{\lambda_1+\lambda_2 \gamma} - \frac{\lambda_2}{\lambda_2+\lambda_1 \gamma} \right)  \nonumber \\ 
   & -\frac{\log(\lambda_1 +\lambda_2 ) }{2\log(2)} - \frac{\lambda_1}{2\log(2)(\lambda_1+\lambda_2 \gamma)} \log(\lambda_1+\lambda_2 \gamma) \nonumber \\
   & - \frac{\lambda_2}{2\log(2)(\lambda_2+\lambda_1 \gamma)} \log(\lambda_2+\lambda_1 \gamma) 
\end{align}
where $m(\gamma)$ is given by eq.(\ref{definitionmgamma}). 
As $m(\gamma)\leq 1$, the slope in $\log(\rho)$ with NOMA-A is then lower than that with NOMA but larger than that with OMA. 

\section{\label{AppendixE}}
 
 From  (\ref{AverageDataRateUserOMA2}), the strong user's asymptotic data rate in OMA is equal to:
  \begin{align} \label{OMA-B-AsymptoticBehavior}
    \mathbb{E}[\tilde{R}_B]& \approx \frac{1}{2 \log(2)} \left( \log(\rho) + \log(2) + \log(\lambda_1+\lambda_2) \right. \nonumber \\
    & \left. - \log(\lambda_1) - \log(\lambda_2) - \gamma_{E}  \right).
\end{align}

In NOMA, the first part $J_{B/\bar{A}}$  tends to $0$ when $\rho >> 1$. The second part $J_{B/A}$ asymptotically tends to:
  \begin{align} \label{RBAsymptoticBehavior}
    \mathbb{E}[R_B] & \approx J_{B/A}  \nonumber \\
      &   \approx  \frac{1}{\log(2)} \log(1+\gamma)\left(\frac{\lambda_1}{\lambda_1+ \lambda_2 \gamma} +\frac{\lambda_2}{\lambda_2+ \lambda_1 \gamma}  \right)   \nonumber \\
      & +  \frac{1}{\log(2)}\frac{1}{(\lambda_1 - \lambda_2)} \left(\lambda_1\log\left(\frac{\lambda_1+ \lambda_2 \gamma}{\lambda_1(1+\gamma)} \right)  \right. \nonumber \\
      & \left. -  \lambda_2\log\left(\frac{\lambda_2+ \lambda_1 \gamma}{\lambda_2(1+\gamma)} \right) \right)
  \end{align}
where we used (\ref{E1Approx}) to approximate function $E_1$.  (\ref{RBAsymptoticBehavior}) shows that $\mathbb{E}[R_B]$  does not depend on $\rho$. Consequently,  $\mathbb{E}[R_B]$  has an asymptote when $\rho  >> 1$.

  Finally, the approximation of $\mathbb{E}[\hat{R}_B]$  when $\rho>>1$ is:
  \begin{align} \label{IbforAsymptoticBehavior}
      \mathbb{E}[\hat{R}_B]  &\approx       J_{B/A} +  \frac{1}{2 \log(2)} \left(\log(\rho)+\log(2)-\gamma_{E}\right) \nonumber \\
      &\left(1 - \frac{\lambda_1}{\lambda_1 + \lambda_2 \gamma} - \frac{\lambda_2}{\lambda_2 + \lambda_1 \gamma}\right)  \nonumber \\
      & -\frac{1}{2\log(2)}\log(\lambda_1 + \lambda_2) + \frac{1}{2\log(2)} \left(\sigma_{1,2}+  \sigma_{2,1} \right)
      \end{align}
      where :
      \begin{align}
          \sigma_{i,j} & =  \frac{\lambda_i}{\lambda_i + \lambda_j \gamma}\log\left(\frac{\lambda_i + \lambda_j \gamma}{2\gamma } \right) + \log\left(\frac{\lambda_i + \lambda_j}{\lambda_j} \right) \nonumber \\
      &+ \log\left(\frac{\lambda_j\gamma}{\lambda_i+\lambda_j\gamma } \right).
      \end{align}
The slope of $ \mathbb{E}[\hat{R}_B]$ in  $\log(\rho)$ is positive but is lower than that of $\mathbb{E}[\tilde{R}_B]$. This completes the proofs. 
  
\bibliographystyle{IEEEbib}


\begin{thebibliography}{10}

\bibitem{ChenVisoz18}
Y.~{Chen}, A.~{Bayesteh}, Y.~{Wu}, B.~{Ren}, S.~{Kang}, S.~{Sun}, Q.~{Xiong},
  C.~{Qian}, B.~{Yu}, Z.~{Ding}, S.~{Wang}, S.~{Han}, X.~{Hou}, H.~{Lin},
  R.~{Visoz}, and R.~{Razavi},
\newblock ``{Toward the Standardization of Non-Orthogonal Multiple Access for
  Next Generation Wireless Networks},''
\newblock {\em IEEE Communications Magazine}, vol. 56, no. 3, pp. 19--27, March
  2018.

\bibitem{DingBhargavaJSAC17}
Z.~{Ding}, X.~{Lei}, G.~K. {Karagiannidis}, R.~{Schober}, J.~{Yuan}, and V.~K.
  {Bhargava},
\newblock ``{A Survey on Non-Orthogonal Multiple Access for 5G Networks:
  Research Challenges and Future Trends},''
\newblock {\em IEEE Journal on Selected Areas in Communications}, vol. 35, no.
  10, pp. 2181--2195, Oct 2017.

\bibitem{DaiComMag15}
L.~{Dai}, B.~{Wang}, Y.~{Yuan}, S.~{Han}, C.~{I}, and Z.~{Wang},
\newblock ``{Non-orthogonal multiple access for 5G: solutions, challenges,
  opportunities, and future research trends},''
\newblock {\em IEEE Communications Magazine}, vol. 53, no. 9, pp. 74--81, Sep.
  2015.

\bibitem{DingPoorComMag17}
Z.~{Ding}, Y.~{Liu}, J.~{Choi}, Q.~{Sun}, M.~{Elkashlan}, C.~{I}, and H.~V.
  {Poor},
\newblock ``{Application of Non-Orthogonal Multiple Access in LTE and 5G
  Networks},''
\newblock {\em IEEE Communications Magazine}, vol. 55, no. 2, pp. 185--191,
  February 2017.

\bibitem{LiuHanzoProceedings17}
Y.~{Liu}, Z.~{Qin}, M.~{Elkashlan}, Z.~{Ding}, A.~{Nallanathan}, and
  L.~{Hanzo},
\newblock ``{Nonorthogonal Multiple Access for 5G and Beyond},''
\newblock {\em Proceedings of the IEEE}, vol. 105, no. 12, pp. 2347--2381, Dec
  2017.

\bibitem{IslamDobreSurvey17}
S.~M.~R. {Islam}, N.~{Avazov}, O.~A. {Dobre}, and K.~{Kwak},
\newblock ``{Power-Domain Non-Orthogonal Multiple Access (NOMA) in 5G Systems:
  Potentials and Challenges},''
\newblock {\em IEEE Communications Surveys Tutorials}, vol. 19, no. 2, pp.
  721--742, Secondquarter 2017.

\bibitem{HossainIEEEAccess16}
M.~S. Ali, H.~Tabassum, and E.~Hossain,
\newblock ``Dynamic user clustering and power allocation for uplink and
  downlink non-orthogonal multiple access (noma) systems,''
\newblock {\em IEEE Access}, vol. 4, pp. 6325--6343, 2016.

\bibitem{LiangNOMA_WCom17}
Y.~{Liang}, X.~{Li}, J.~{Zhang}, and Z.~{Ding},
\newblock ``{Non-Orthogonal Random Access for 5G Networks},''
\newblock {\em IEEE Transactions on Wireless Communications}, vol. 16, no. 7,
  pp. 4817--4831, July 2017.

\bibitem{ShirDohlerJSAC17}
M.~{Shirvanimoghaddam}, M.~{Condoluci}, M.~{Dohler}, and S.~J. {Johnson},
\newblock ``{On the Fundamental Limits of Random Non-Orthogonal Multiple Access
  in Cellular Massive IoT},''
\newblock {\em IEEE Journal on Selected Areas in Communications}, vol. 35, no.
  10, pp. 2238--2252, Oct 2017.

\bibitem{AbbasVuceticTCom19}
R.~{Abbas}, M.~{Shirvanimoghaddam}, Y.~{Li}, and B.~{Vucetic},
\newblock ``{A Novel Analytical Framework for Massive Grant-Free NOMA},''
\newblock {\em IEEE Transactions on Communications}, vol. 67, no. 3, pp.
  2436--2449, March 2019.

\bibitem{WangIoT19}
H.~{Wang} and A.~O. {Fapojuwo},
\newblock ``{Design and Performance Evaluation of Successive Interference
  Cancellation-Based Pure Aloha for Internet-of-Things Networks},''
\newblock {\em IEEE Internet of Things Journal}, vol. 6, no. 4, pp. 6578--6592,
  Aug 2019.

\bibitem{EEMaximization_Musavian_TWC15}
L.~{Musavian} and Q.~{Ni},
\newblock ``{Effective Capacity Maximization With Statistical Delay and
  Effective Energy Efficiency Requirements},''
\newblock {\em IEEE Trans. Wireless Commun.}, vol. 14, no. 7, pp. 3824--3835,
  July 2015.

\bibitem{Tradeoff_Yu_Musavian_TWC16}
W.~{Yu}, L.~{Musavian}, and Q.~{Ni},
\newblock ``{Tradeoff Analysis and Joint Optimization of Link-Layer Energy
  Efficiency and Effective Capacity Toward Green Communications},''
\newblock {\em IEEE Trans. Wireless Commun.}, vol. 15, no. 5, pp. 3339--3353,
  May 2016.

\bibitem{BelloChorti2020}
B.~{Mouktar}, W.~{Yu}, A.~{Chorti}, and L.~{Musavian},
\newblock ``{Performance Analysis of NOMA Uplink Networks under Statistical QoS
  Delay Constraints},''
\newblock in {\em to appear in IEEE ICC 2020}, June 2020, pp. 1--7.

\bibitem{NOMAPFVTC14}
X.~Chen, A.~Benjebbour, A.~Li, and A.~Harada,
\newblock ``Multi-user proportional fair scheduling for uplink non-orthogonal
  multiple access (noma),''
\newblock in {\em 2014 IEEE 79th Vehicular Technology Conference (VTC Spring)},
  May 2014, pp. 1--5.

\bibitem{PFNOMAVTC16}
F.~Liu, P.~Mähönen, and M.~Petrova,
\newblock ``Proportional fairness-based user pairing and power allocation for
  non-orthogonal multiple access,''
\newblock in {\em 2015 IEEE 26th Annual International Symposium on Personal,
  Indoor, and Mobile Radio Communications (PIMRC)}, Aug 2015, pp. 1127--1131.

\bibitem{PischellaNOMA2019WCL}
M.~{Pischella} and D.~{Le Ruyet},
\newblock ``{NOMA-Relevant Clustering and Resource Allocation for Proportional
  Fair Uplink Communications},''
\newblock {\em IEEE Wireless Commun. Lett.}, vol. 8, no. 3, pp. 873--876, June
  2019.

\bibitem{SedaghatTWC18}
M.~A. Sedaghat and R.~R. M{\"u}ller,
\newblock ``On user pairing in uplink noma,''
\newblock {\em IEEE Transactions on Wireless Communications}, vol. 17, no. 5,
  pp. 3474--3486, May 2018.

\bibitem{CelikGlobecom17}
A.~Celik, R.~M. Radaydeh, F.~S. Al-Qahtani, A.~H.~A. El-Malek, and M.~S.
  Alouini,
\newblock ``Resource allocation and cluster formation for imperfect noma in
  dl/ul decoupled hetnets,''
\newblock in {\em 2017 IEEE Globecom Workshops (GC Wkshps)}, Dec 2017, pp.
  1--6.

\bibitem{FairPAOviedoTVT18}
J.~A. Oviedo and H.~R. Sadjadpour,
\newblock ``A fair power allocation approach to noma in multiuser siso
  systems,''
\newblock {\em IEEE Transactions on Vehicular Technology}, vol. 66, no. 9, pp.
  7974--7985, Sept 2017.

\bibitem{ZengPoor19}
M.~{Zeng}, A.~{Yadav}, O.~A. {Dobre}, and H.~V. {Poor},
\newblock ``{Energy-Efficient Joint User-RB Association and Power Allocation
  for Uplink Hybrid NOMA-OMA},''
\newblock {\em IEEE Internet of Things Journal}, vol. 6, no. 3, pp. 5119--5131,
  June 2019.

\bibitem{LeiNOMA16}
L.~{Lei}, D.~{Yuan}, C.~K. {Ho}, and S.~{Sun},
\newblock ``{Power and Channel Allocation for Non-Orthogonal Multiple Access in
  5G Systems: Tractability and Computation},''
\newblock {\em IEEE Transactions on Wireless Communications}, vol. 15, no. 12,
  pp. 8580--8594, Dec 2016.

\bibitem{TweedIEEEAccess17}
D.~{Tweed}, M.~{Derakhshani}, S.~{Parsaeefard}, and T.~{Le-Ngoc},
\newblock ``{Outage-Constrained Resource Allocation in Uplink NOMA for Critical
  Applications},''
\newblock {\em IEEE Access}, vol. 5, pp. 27636--27648, 2017.

\bibitem{ZhaiAdmissionCULNOMA18}
D.~{Zhai} and R.~{Zhang},
\newblock ``{Admission Control and Resource Allocation for Multi-Carrier Uplink
  NOMA Networks},''
\newblock {\em IEEE Wireless Communications Letters}, vol. 7, no. 6, pp.
  922--925, Dec 2018.

\bibitem{XuSchonerOutage16}
P.~{Xu}, Y.~{Yuan}, Z.~{Ding}, X.~{Dai}, and R.~{Schober},
\newblock ``{On the Outage Performance of Non-Orthogonal Multiple Access With
  1-bit Feedback},''
\newblock {\em IEEE Transactions on Wireless Communications}, vol. 15, no. 10,
  pp. 6716--6730, Oct 2016.

\bibitem{DingPoorSigProcLetters14}
Z.~{Ding}, Z.~{Yang}, P.~{Fan}, and H.~V. {Poor},
\newblock ``{On the Performance of Non-Orthogonal Multiple Access in 5G Systems
  with Randomly Deployed Users},''
\newblock {\em IEEE Signal Processing Letters}, vol. 21, no. 12, pp.
  1501--1505, Dec 2014.

\bibitem{LiuOutageNOMA18}
Y.~{Liu}, M.~{Derakhshani}, and S.~{Lambotharan},
\newblock ``{Outage Analysis and Power Allocation in Uplink Non-Orthogonal
  Multiple Access Systems},''
\newblock {\em IEEE Communications Letters}, vol. 22, no. 2, pp. 336--339, Feb
  2018.

\bibitem{BelloVTCArxiv20}
M.~Bello, W.~Yu, M.~Pischella, A.~Chorti, and I.~Fijalkow,
\newblock ``{Flexible Multiple Access Enabling Low-Latency
  Communications:Introducing NOMA-R},''
\newblock {\em arXiv:2001.10637}, Jan. 2020.

\bibitem{TableIntegralsEI69}
M.~{Geller} and E.W. {Ng},
\newblock ``{A Table of Integrals of the Exponential Integral},''
\newblock {\em Journal of Research of the National Bureau of Standards}, vol.
  73B, no. 3, pp. 191--210, Sep 1969.

\bibitem{HandbookMathematicalFunctions}
W.~{Gautschi} and W.F {Cahill},
\newblock ``{Handbook of Mathematical Functions with Formulas, Graphs and
  Mathematical Tables, chap. 5, Exponential Integral and Related Functions},''
\newblock {\em Journal of Research of the National Bureau of Standards}, vol.
  55, no. 5, pp. 227--253, June 1964.

\bibitem{yang_alouini_2011}
H.-C. Yang and M.-S. Alouini,
\newblock {\em Order Statistics in Wireless Communications: Diversity,
  Adaptation, and Scheduling in MIMO and OFDM Systems},
\newblock Cambridge University Press, 2011.

\bibitem{KoAlouiniMRC07}
Y.~{Ko}, H.~{Yang}, S.~{Eom}, and M.~{Alouini},
\newblock ``{Adaptive Modulation with Diversity Combining Based on
  Output-Threshold MRC},''
\newblock {\em IEEE Trans. Wir. Commun.}, vol. 6, no. 10, pp. 3728--3737,
  October 2007.

\end{thebibliography}

\end{document}